\documentclass[12pt]{article}
\usepackage{epsf,afterpage}
\title{ Triangular and $Y$-shaped hadrons in QCD}
\author{ D.S.Kuzmenko and Yu.A.Simonov\\
{\it State Research Center}\\{\it Institute of Theoretical and
Experimental Physics,} \\{\it  117218 Moscow, Russia}}
 \date{}
\newcommand{\be}{\begin{equation}}
 \newcommand{\ee}{\end{equation}}
 \def\la{\mathrel{\mathpalette\fun <}}
\def\ga{\mathrel{\mathpalette\fun >}}
\def\fun#1#2{\lower3.6pt\vbox{\baselineskip0pt\lineskip.9pt
\ialign{$\mathsurround=0pt#1\hfil ##\hfil$\crcr#2\crcr\sim\crcr}}}
\newcommand{\bc} {\begin{center}}
\newcommand{\ec} {\end{center}}
\newcommand{\lan}{\langle}
\newcommand{\ran}{\rangle}
\newcommand{\tr}{\mathrm{tr}}
\newcommand{\mr}[1]{\mathrm{#1}}
\begin{document}
\maketitle

\begin{abstract}
Gauge invariant extended configurations
are considered for the three fundamental (quarks) or adjoint (gluons)
particles. For quarks it is shown that the  $Y$-shaped
 configuration is the only possible. For adjoint
sources both  the $Y$-shaped  and triangular configurations may
realize. The  corresponding static potentials  are calculated in the
Method of Field Correlators and in the case of baryon shown to be
consistent with the lattice simulations. For adjoint sources the
 potentials of $Y$-shaped and $\Delta$-shaped configurations turn out
to be close to each other, which leads to almost degenerate masses of
$3^{--}$ 3g glueballs and odderon trajectories.
\end{abstract}

1. To make conclusions on the structure of gluonic fluxes confining
color charges in physical states, one has to start from the
considering of the space-extended gauge invariant wave function of
the hadron.
It is easy then to show that in the case of the static charges the
Green function of the hadron reduces to the Wilson loop, which in the
case of  the baryon  has the $Y$-type shape and consists from
the three contours formed by the quark trajectories and joined at the
point of the string junction \cite{1,2}.  The Wilson loop of
the tree adjoint sources is less known than the 3q one. In the
 paper we will show that it can have both $Y$-type and
$\Delta$-type shape.

Using the formalism of the Method of the Field Correlators (MFC),
we compute the static potentials, corresponding to the
Wilson loops of the hadrons. In the case of the baryon the static
potential was used long ago in many dynamical calculations \cite{3,4}.
Recently it has been computed in lattice gauge theory in a number of
papers \cite{7,8,9}. We show in the paper that our potential is in a
full agreement with the lattice studies.

In the case of adjoint sources we find that the $Y$-type and
$\Delta$-type potentials remain near each other at the characteristic
hadronic size.  Using that we estimate masses of lowest $3g$
glueballs, lying on the corresponding odderon trajectories, and show
that they are close to each other, implying that there are two
possible odderon trajectories with not much different Regge
slopes.  A short discussion of physical implications of these results
concludes the paper.

To avoid confusion, we
should stress that the term "$\Delta$ configuration"
used in \cite{6,7,8,9} in the context of the static
baryon potential, refers to the perimeter behaviour of the potential
and not to the gauge invariant configurations as well as to the
structures of fluxes discussed in the present paper.

2. Hadron building in SU(3) starts with listing elementary
building blocks: quarks $q^\alpha,~\alpha=1,2,3$, gluons (or
adjoint static sources) $g^a,~ a=1,...8$, parallel transporters
(PT) in fundamental representation  $\Phi_\alpha^\beta(x,y)=
(P\exp ig \int A_\mu(z) dz_\mu)_\alpha^\beta$, adjoint parallel
transporters $\Phi_{ab} (x,y)$, generators $t^{(a)\beta}_\alpha$
symmetric symbols $\delta_\alpha^\beta,~ \delta_{ab},~ d^{abc}$, and
antisymmetric ones, $e_{\alpha\beta\gamma}$ and $f^{abc}$. Note
that we always use Greek indices for fundamental representation
and Latin ones for the adjoint.

To construct a real extended (not point-like) hadron one uses all
listed elements, PT included, and forms a white (gauge-invariant)
combination. It is convenient to form an extended quark
(antiquark) operator
$$
q^\alpha(x,Y)\equiv q^\beta(x) \Phi^\alpha_\beta(x,Y);
$$
\be
\bar q_\alpha(x,Y)=\bar q_\beta(x)\Phi^\beta_{\alpha} (x,Y).
\label{1}
\ee
In this way
one has for the Y-shaped baryon:
\be B_Y(x,y,z,Y) =
e_{\alpha\beta\gamma}q^\alpha(x,Y)
q^\beta(y,Y)q^\gamma(z,Y).
\label{2}
\ee
One can also define quark
operator with two lower indices:
$e_{\alpha\beta\gamma}q^{\alpha}(x)\equiv q_{\beta\gamma}(x)$.
However an attempt
 to create a gauge-invariant combination from 3 operators
 $q_{\beta\gamma}(x)$ and  3 PT to construct a $\Delta$-type
 configuration fails: the structure
 \be
 B_\Delta(x,y,z) = q_{\alpha\beta} (x) \Phi^\beta_\gamma(x,y)
 q_{\gamma\delta}(y) \Phi^\delta_\varepsilon(y,z)
 q_{\varepsilon\rho}(z) \Phi^\rho_\alpha(z,x)
 \label{3}
 \ee
is not gauge invariant, that can be checked directly, substituting in
(\ref{3}) $q^\alpha(x)\to U^\alpha_\beta(x)q^\beta(x)$. One can try
all combinations, but it is impossible  to form a continuous chain of
indices to represent the $\Delta$-type structure using as operators
$q_\alpha$ as $q_{\alpha\beta}$. Thus one can conclude
that the $Y$-shaped configuration is the only possible gauge-invariant
configuration of wave function for baryons.

One may wonder, what is the relation between the spacial structures of
wave function of hadrons and their gluonic flux? The answer
from the flux-tube models is that these structures coincide.
In realistic lattice calculations one is to use the Wilson loop,
which describes the gauge invariant state of hadron generated at some
initial and annihilated at final moment of time. As is well known, for
static charges the Wilson loop  consists of two wave functions
considered above, joined by parallel transporters. Let us imagine now
some evolution of the fluxes  in baryon which would lead to the
emerging of the $\Delta$-type configuration of fluxes in some
intermediate time. First of all we should note that the cross-section
of the Wilson loop by the time-like hypersurface will recover a
gauge-invariant 3q state, i.e. the $Y$-type configuration. To have a
$\Delta$-shape for fluxes one should admit that the fluxes have no
relation with the wave function, which is unprobable.

Consider now the adjoint source $g^a(x) t^{(a)\beta}_\alpha \equiv
G^\beta_\alpha(x)$. We do not specify here the Lorentz structure of
$g^a(x)$, but only impose condition that it should gauge transform
homogeneously, $G_\alpha^\beta\to
U^{+\beta}_{{}\beta'}G_{\alpha'}^{\beta'}U^{\alpha'}_{\alpha}$.
Therefore $g^a(x)$ can be either the field strength $F^a_{\mu\nu}(x)$,
or valence gluon field $a^a_\mu(x)$ in the background-field
perturbation theory \cite{BPTh}. It is easy to construct a
$\Delta$-type configuration for 3 such sources;
\be
G_\Delta(x,y,z) =G_\alpha^\beta (x) \Phi_\beta^\gamma(x,y)
G_\gamma^\delta(y) \Phi_\delta^\varepsilon (y,z) G_\varepsilon
^\rho(z) \Phi_\rho^\alpha(z,x).
\label{4}
\ee
 It is clear that in
(\ref{4}) all repeated indices form gauge-invariant combinations,
and $G_\Delta (x,y,z)$ is a gauge-invariant $\Delta$-type
configuration, which was used previously for the $3g$ glueball in
\cite{11}.

But one can persuade oneself that (\ref{4}) is not the only $3g$
gauge-invariant configuration. Consider adjoint sources and
adjoint PT (here distinguishing upper and lower indices is not
necessary) and form as in (\ref{1}) an extended gluon operator:
\be g_a(x,Y)\equiv g^b(x) \Phi_{ab} (x,Y)
\label{5}
\ee
 and an $Y$-shaped configuration
 \be
 G^{(f)}_Y(x,y,z,Y)=f^{abc}g_a(x,Y) g_b(y,Y) g_c(z,Y).
\label{6}
 \ee
 In the same way one constructs $G^{(d)}_Y$replacing $f$ by $d$ in 
 (\ref{6}).It is clear that $G_Y$ is gauge-invariant and should be
 considered on the same grounds as $G_\Delta$.

\begin{figure}[!t]
\epsfxsize=8cm
\hspace*{4.35cm}
\epsfbox{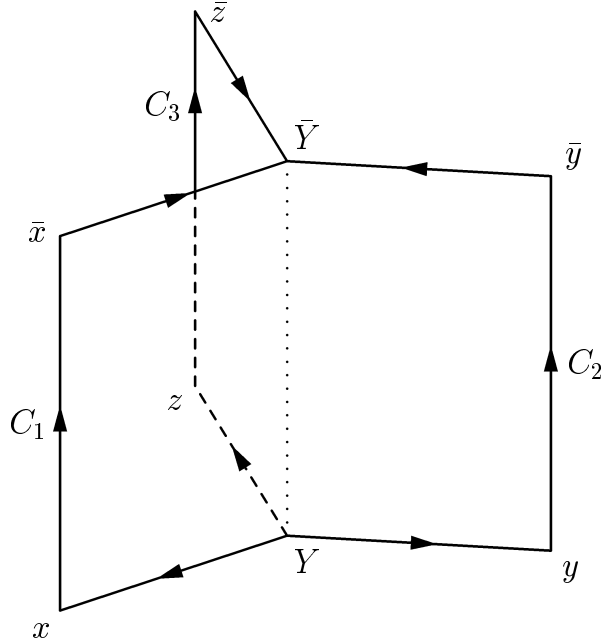}
\caption{$Y$-shaped Wilson loop}
\label{YWloop}
\end{figure}

 At this point it is necessary to clarify how (\ref{2}), (\ref{6})
 generate Green's functions and Wilson loops.

 To this end consider initial and final states made of (\ref{2}),
 (\ref{4}), (\ref{6}) and for simplicity of arguments take all
 fundamental and  adjoint sources  to be static,
 i.e. propagating only in Euclidean time.

 Then the Green's function for the object will be
 \be
{\cal G}_i(\bar X, X)=\lan \Psi^+ _i(\bar X)\Psi_i(X)\ran
\label{7}
\ee
 where $\Psi_i=G_\Delta, G_Y, B_Y;~X=x,y,z$ for $G_\Delta$ and
 $x,y,z,Y$
otherwise. Now it is important that vacuum average in (\ref{7})
produces a product of Green's functions for quarks or for valence
gluons in the external vacuum gluonic field, which is proportional to
the corresponding PT, fundamental -- for quarks and adjoint -- for
gluons. Namely,  $$ \lan \bar q_\beta (\bar x) q^\alpha(x)\ran\sim
\Phi_\beta^\alpha (\bar x, x),
$$
\be
 \lan  g^a
(\bar x) g^b(x)\ran\sim \Phi_{ab}(\bar x, x).
\label{8}
\ee
(This
statement is well known for static sources, for relativistic
quarks and gluons this follows directly from the exact
Fock-Feynman-Schwinger representation (FFSR), see \cite{12,13} and for
a review {\cite{14}).

As a result one obtains a gauge-invariant Wilson-loop combination
for each Green's function (\ref{7}). In particular for $B_Y$
(\ref{2})
 one has a familiar 3-lobe Wilson loop $W_Y$:
 \be
 W_Y(\bar X, X)=\tr_Y\prod^3_{i=1} W_i(C_i),
\label{9}
 \ee
 where $\tr_Y=\frac16 e_{\alpha\beta\gamma} e_{\alpha'\beta'
 \gamma'}$, and the contour $C_i$ in the open loop $W_i$ passes
 from $Y$ to $\bar Y$ through points $x,\bar x\,(i=1)$,\quad  $y,\bar
y  \,(i=2)$, or  $z, \bar z\, (i=3)$, as shown in Fig.\ref{YWloop}.

\begin{figure}[!t]
\epsfxsize=8cm
\hspace*{4.35cm}
\epsfbox{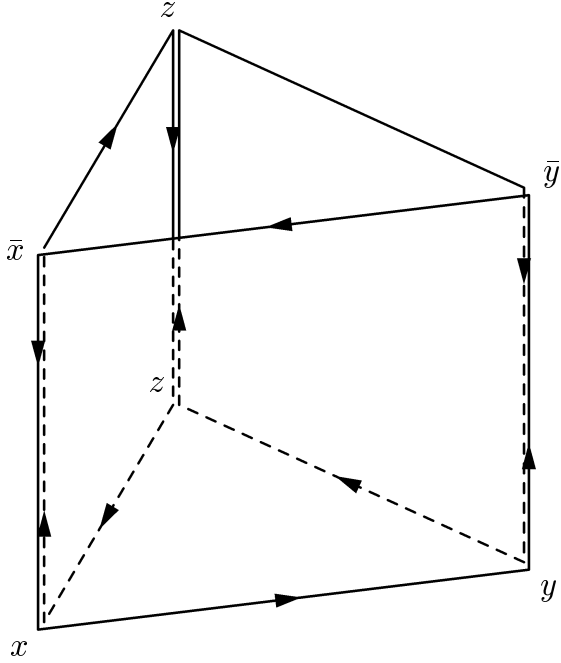}
\caption{$\Delta$-shaped Wilson loop}
\label{DeltaWloop}
\end{figure}

 This situation is well-known and was exploited in numerous
 applications. Relatively less known are the Wilson-loop
 configurations for $G_Y$ and $G_\Delta$. In the first case the
 structure is the same with the replacement of fundamental lines
 and symbols by the adjoint ones: $e_{\alpha\beta\gamma}\to
 f^{abc}$ or $e_{\alpha\beta\gamma}\to d^{abc}$, $\Phi_\alpha^\beta
\to \Phi_{ab}$, so that the whole  structure in (\ref{9}) is the same
with this replacement. Contrary to the baryon case, we can contract
adjoint indices in two ways, using antisymmetric symbol $f^{abc}$
or symmetric one $d^{abc}$. The proper choice is related to
the Bose-statistics of the gluon system which ensures  the full
coordinate-spin function to be symmetric.

In the case of $G_\Delta$ using (\ref{4}) and (\ref{8}) one can
write the resulting structure symbolically as follows
\be
{\cal G}_\Delta(\bar X, X)= \Delta_{a'b'c'}(\bar x, \bar y, \bar z)
\Phi_{a'a} (\bar x, x) \Phi_{b'b} (\bar y, y) \Phi_{c'c} (\bar z,
z) \Delta_{abc} (x,y,z)
\label{10}
\ee
where we have denoted
\be
\Delta_{abc} (x,y,z) =t^{(a)\beta}_\alpha \Phi_\beta^\gamma(x,y)
t^{(b)\delta}_{\gamma} \Phi^\varepsilon_\delta(y,z)
t^{(c)\rho}_\varepsilon \Phi^\alpha_\rho(z,x) .
\label{11}
\ee
To
understand better the structure of (\ref{10}), one can use the
large $N_c$ approximation, in which case one has
\be
\Psi^{\beta\beta'}_{\alpha\alpha'} \equiv t^{(a')\beta'}_{\alpha'}
\Phi_{a'a} (\bar x, x)
t^{(a)\beta}_\alpha\approx\frac12\Phi_\alpha^{\beta'} (x,\bar x)
\Phi_{\alpha'}^\beta(\bar x, x).
\label{12}
\ee

As a
result in this approximation ${\cal G}_\Delta$ appears to be a product
of 3 fundamental closed loops, properly oriented with respect to each
other
\be
{\cal G}_\Delta (\bar X, X)\sim W(\bar x,\bar y|x,y) W(\bar y, \bar
z|y,z) W(\bar z, \bar x|z,x)\equiv W_{\Delta}(\bar X,X),
\label{13}
\ee
it is displayed in Fig. \ref{DeltaWloop}.

3. Static potentials for configurations (\ref{2}), (\ref{4}),
(\ref{6}) can be computed using Field Correlator Method (FCM)
\cite{15}, through the equation
\be
V=- \lim_{T\to \infty} \frac{1}{T}\ln \langle W\rangle,
\label{VW}
\ee
where $T$ is the time extension of the Wilson loop.

\begin{figure}[!t]
\epsfxsize=12cm
\vspace{-3mm}
\hspace*{2.35cm}
\epsfbox{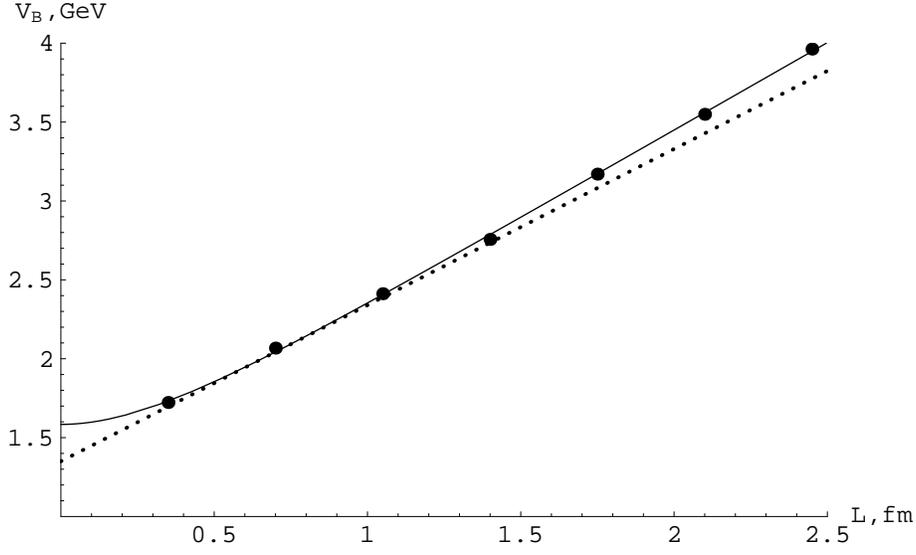}
\caption{The lattice nonperturbative baryon potential
from \cite{9} (points) for lattice parameter $\beta=5.8$ and  MFC
potential $V^{(B)}$ (solid line) with parameters   $\sigma=0.22$
GeV$^2$ and  $T_g=0.12$ fm vs. the minimal length of the string $L$.
The dotted line is a tangent at $L=0.7$ fm. The figure is taken from
\cite{K}.}
\label{fig3}
\end{figure}

For the baryon in the case of 3
quarks at the vertices of an equilateral triangle, at the distance
$R$ from the string junction Y, the static baryon
potential reads as \cite{K}
\be
   V^{(B)}(R)=3V^{(M)}(R)+V^{(\mathrm{nd})}(R),
\label{14}
\ee
where
\be
V^{(M)}(R)=\frac{2\sigma}{\pi}\left\{
R\int_0^{R/T_g}d\,x\, x\, K_1(x)-T_g
\left(2-\frac{R^2}{T_g^2}K_2\left(\frac R{T_g}\right)\right)\right\}
\label{VM}
\ee
 is the mesonic confining  potential with the
asymptotic slope $\sigma\approx 0.18$ GeV$^2$ and the gluonic
correlation length $T_g=0.12\div 0.2$ fm \cite{Tg}, and the
nondiagonal part of the potential,
\be
V^{\mathrm{(nd)}}(R)=\frac2{\sqrt{3}}\sigma T_g-
\frac{3\sqrt{3}}{2\pi}\frac{\sigma R^2}{T_g}
\int_{\frac{\pi}6}^{\frac{\pi}3}\frac{d\,\varphi}{\cos \varphi}
 K_2\left(\frac{\sqrt{3}R}{2T_g\cos \varphi}\right),
\label{Vnd}
\ee
appears due to the
interference of the gluonic fields on different lobes of the Wilson
loop. Note the difference in the overall factor $-1/2$ with
the previous calculations \cite{10}, where it was erroneously
omitted. Let us denote $L\equiv 3R$ the total length of the string. In
Fig. 3 from \cite{K} the dependence of lattice nonperturbative baryon
potential from \cite{9} on $L$ along with the MFC potential
(\ref{14})-(\ref{Vnd}) is shown. One can see that our potential is in
the complete agreement with the lattice results.
In the asymptotic region $L\ga 1.5$ fm the potential has a linear form
\be
V^{(B)}(R)\approx \sigma L +
\left(\frac2{\sqrt{3}}-\frac{12}{\pi}\right)\sigma T_g.
\label{14c}
\ee
The dotted tangent in Fig. 3 demonstrates that in the range 0.3 fm$\la
L\la 1.5$ fm the lattice data can be described by the linear potential
with the slope some 10\% less than $\sigma$.

The potential written  so far contains only the nonperturbative
confining part. To obtain the total potential we should  add to
it the perturbative color-coulombic potential
\be
V^{\mathrm{pert}}_{\mathrm{(fund)}}(r)=-\frac32\,
\frac{C_2(\mathrm{fund})\alpha_s}{r},
\label{14(d)}
\ee
where $r=\sqrt{3}R$ is the interquark distance in the equilateral
triangle and $C_2(\mathrm{fund})=4/3$.

\begin{figure}[!t]
\epsfxsize=12cm
\vspace{-3mm}
\hspace*{2.35cm}
\epsfbox{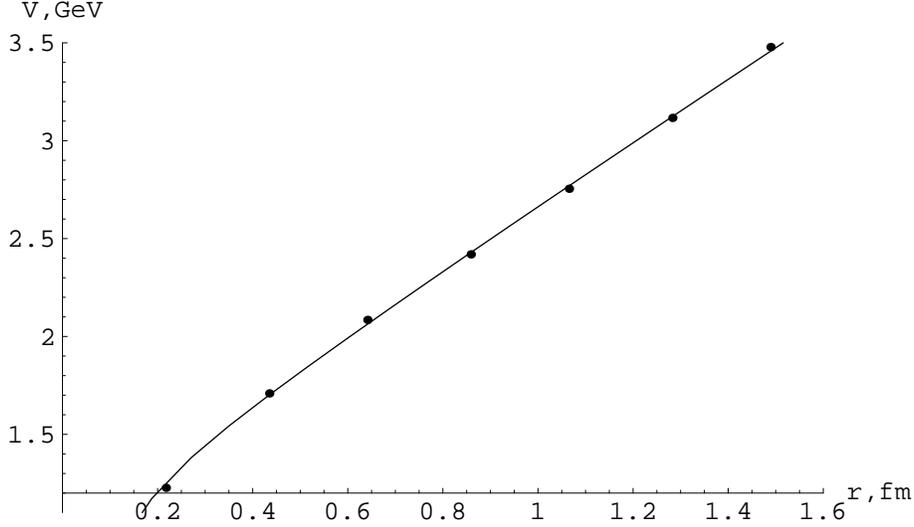}
\caption{The lattice baryon potential in the equilateral triangle with
quark separations $r$ from \cite{8} (points) at $\beta=5.8$ and the
MFC potential $V^{(B)}+V^{\mathrm{pert}}_{\mathrm{(fund)}}$ (solid
line) at  $\alpha_s=0.18$, $\sigma=0.18$ GeV$^2$, and $T_g=0.12$ fm.}
\label{fig4}
\end{figure}

In Fig. \ref{fig4} lattice data from the last ref. of
 \cite{8}  and the potential
$V_B(r)+V^{\mr{pert}}_{\mr{(fund)}}(r)$  are shown.
One can see that our results are in the complete agreement with this
independent set of the lattice data as well.

In a similar way one can write the static potential for the adjoint
sources, neglecting the nondiagonal term, which is different in
symmetric and antisymmetric states:
\be
V_Y^{(G)}(R)=\frac{C_2(\mathrm{adj})}{C_2(\mathrm{fund})}V_Y^{(B)}(R)=
\frac94 V_Y^{(B)}(R).
\label{15}
\ee

\begin{figure}[!t]
\epsfxsize=12cm
\vspace{-3mm}
\hspace*{2.35cm}
\epsfbox{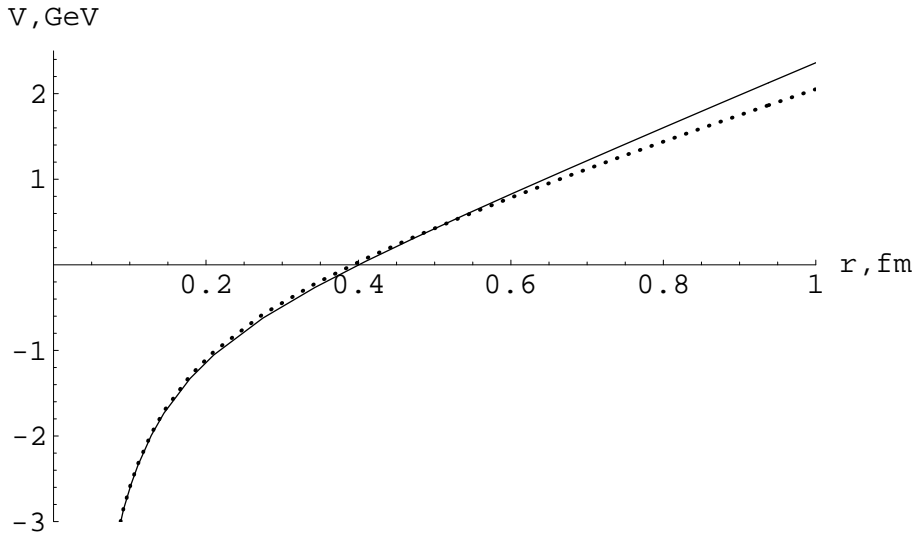}
\caption{ Glueball potentials
$V_Y^{(G)}(r)+V^{\mathrm{pert}}_{\mathrm{(adj)}}(r)$  (solid curve)
and  $V_\Delta^{(G)}(r)+ V^{\mathrm{pert}}_{\mathrm{(adj)}}(r)$
(dashed curve) in equilateral triangle with the quark separations $r$
for $\alpha_s=0.3$, $\sigma$=0.18 GeV$^2$ and $T_g=0.12$ fm. The
nondiagonal terms are neglected.}
\label{YDelta}
\end{figure}

Consider the $\Delta$-configuration
in the approximation (\ref{12}). In this case
$V_\Delta^{(G)}(R)$ reduces to the sum of the mesonic potentials
corresponding to area laws for all three loops minus the
nondiagonal interference term, and one obtains just as in \cite{10}
\be
   V_\Delta^{(G)}(r)=3V^{(M)}(r)+V^{(\mathrm{nd})}(r).
\label{16}
\ee

Along with the adjoint perturbative potential
\be
V^{\mathrm{pert}}_{\mathrm{(adj)}}(r)=-\frac32\,
\frac{C_2(\mathrm{adj})\alpha_s}{r},
\label{17}
\ee
where $C_2(\mathrm{adj})=3$, we plot both $V_Y^{(G)}$ and
$V_\Delta^{(G)}$  in Fig. \ref{YDelta} without the
interference terms. We see from the figure that the curves intersect
at $r\approx 0.5$ fm, and are very close to each other.

4. To summarize our results, we have considered possible
gauge-invariant configurations of 3 fundamental or adjoint sources
and corresponding Wilson loops, which have $Y$-type shape for
fundamental charges and may have both $Y$-type and $\Delta$-type
shapes for the adjoint ones.  We have shown that the static baryon
potential obtained in the MFC is in the complete agreement with the
lattice data.

For adjoint sources it was demonstrated that two
possible configurations yield static potentials differing only a
little. This in turn implies that 3-gluon glueballs \cite{11} may be
of two distinct types, with no direct transitions between them
(quark-containing hadrons must be involved as intermediate states).
The mass of the $\Delta$-shaped $3^{--}$ glueball was found in
\cite{11} to be $M_\Delta^{(3g)}=3.51$ GeV for $\sigma_f=0.18$
GeV$^2$ (or 4.03 GeV for $\sigma_f=0.238$ GeV$^2$ to be compared with
lattice one calculated in \cite{16} 4.13$\pm$0.29 GeV). The mass of
the $Y$-shaped glueball can easily be computed from the baryon mass
calculated in \cite{17}, multiplying it by $\sqrt{9/4}=3/2$. In this
way one obtains $M_Y^{(3g)}=3.47$ GeV ($\sigma_f=0.18$ GeV$^2$). The
slope of the corresponding odderon trajectory is almost the same and
corresponds to $g-gg$-configuration. Thus one obtains the
$\Delta$-odderon (slope)$^{-1}$ to be twice the standard Regge slope,
while for $Y$-odderon it is $\frac94$ of the standard slope. In both
cases the intercept comes out as in \cite{11} to be rather low ($-1.8$
for the $Y$-shape and $-2.4$ for the $\Delta$-shape) implying very
small odderon contribution to reactions under investigation \cite{18}
in agreement with measurements. We plan to perform more accurate
calculations of glueball potentials and spectra  taking into account
the string-string interference in subsequent publications.

Both authors acknowledge
partial support from grants RFFI 00-15-96786, 00-02-17836 and INTAS
00-00110, 00-00366. One of the authors (Yu.S.) is grateful for useful
discussions to D. Richards and R. Edwards; he was supported by DOE
contract DE-AC05-84ER40150 under which SURA operates the Thomas
Jefferson National Accelerator Facility.


\begin{thebibliography}{99}
\looseness=-1


\bibitem{1} X. Artru, Nucl.Phys. {\bf B85}, 442 (1975).

\bibitem{2} H.G. Dosch and V. Mueller, Nucl.Phys. {\bf B116}, 470
(1976).

\bibitem{3} J. Carlson, J. Kogut, and V.R. Pandharipande,
Phys.Rev. {\bf D27}, 233 (1983);\\
N. Isgur and J. Paton, Phys.Rev. {\bf D31}, 2910 (1985).

\bibitem{4} Yu.A. Simonov, Phys. Lett. {\bf B228}, 413 (1989),ibid.
{\bf B515}, 137 (2001);\\
M. Fabre de la Ripelle and Yu.A. Simonov, Ann.Phys.(N.Y.) {\bf 212},
235 (1991).

\bibitem{7} G.S. Bali, Phys.Rept. {\bf 343}, 1 (2001).

\bibitem{8} C. Alexandrou, Ph. de Forcrand, and A.Tsapalis,
Phys.Rev. {\bf D65}, 054503 (2002);\\
 C. Alexandrou, Ph. de Forcrand, and O. Jahn, hep-lat/0209062.

\bibitem{9} T.T. Takahashi {\it et al.}, Phys.Rev. D {\bf 65},
114509 (2002).

\bibitem{6} J.M. Cornwall, Phys.Rev. {\bf D54}, 6527 (1996).

\bibitem{BPTh}
B.S. De Witt, Phys.Rev. {\bf 162}, 1195, 1239 (1967);\\
L.F.Abbot, Nucl.Phys. {\bf B185}, 189 (1981);\\
Yu.A. Simonov, Phys.At.Nucl. {\bf 58}, 107 (1995), hep-ph/9909237.

\bibitem{11} A.B. Kaidalov and Yu.A. Simonov, Phys.Atom.Nucl. {\bf
63}, 1428 (2000);\\
 Phys. Lett. {\bf B477}, 163 (2000).

\bibitem{12} R.P. Feynman, Phys. Rev. {\bf 80} 440 (1950); ibid.
 {\bf 84} 108 (1951);\\
V.A. Fock, Izvestya Akad. Nauk USSR, OMEN, 1937, p.557;\\
J. Schwinger, Phys. Rev. {\bf 82} 664 (1951).

\bibitem{13} Yu.A. Simonov, Nucl. Phys. {\bf B307}, 512 (1988);\\
Yu.A. Simonov and J.A. Tjon, Ann.Phys. {\bf 228}, 1 (1993).

\bibitem{14} Yu.A. Simonov and J.A. Tjon, in the Michael  Marinov
Memorial Volume, "Multiple facets of quantization and supersymmetry",
Eds. M. Olshanetsky and A. Vainshtein (World
Scientific), hep-ph/0201005.

\bibitem{15} H.G. Dosch and Yu.A. Simonov, Phys. Lett. {\bf
B205}, 339 (1988).

\bibitem{K}
D.S. Kuzmenko, to appear in Yad.Fiz., hep-ph/0204250.

\bibitem{Tg}
 M. Campostrini, A. Di Giacomo and G. Mussardo, Z.Phys. C {\bf 25},
173    (1984);\\
A. Di Giacomo and H. Panagopoulos, Phys.Lett. B {\bf 285 }, 133
(1992);\\ A. Di Giacomo, E. Meggiolaro and H. Panagopoulos, Nucl.Phys.
 B   {\bf 483 }, 371 (1997);\\
G.S. Bali, N. Brambilla and A. Vairo, Phys.Lett. {\bf B421}, 265
(1998);\\
Yu.A. Simonov, Nucl.Phys. {\bf B592}, 350 (2001).

\bibitem{10} D.S. Kuzmenko and Yu.A. Simonov, Phys.Lett. {\bf B494}
81 (2000); Phys.At.Nucl. {\bf 64} 107 (2001).

\bibitem{interf}
Yu.A.Simonov, Phys.Rept. {\bf 320}, 265 (1999).


\bibitem{16} C. Morningstar, M. Peardon,  Nucl.Phys.Proc.Suppl.
{\bf 73},  927 (1999);
Phys.Rev. {\bf D60}, 034509 (1999).

\bibitem{17}  B.O. Kerbikov and Yu.A.Simonov, Phys.Rev. {\bf D62},
093016 (2000).

\bibitem{18} J. Olsson et. al. (H1 collaboration), hep-ex/0112012.


\end{thebibliography}
\end{document}